\documentstyle[twocolumn]{article}
\begin{document}
\twocolumn[\hsize\textwidth\columnwidth\hsize\csname
@twocolumnfalse\endcsname

\def\hpione{{h_\pi^1}}
\def\hpieff{{h_\pi^\sst{EFF}}}
\def\gpi{{g_\pi}}
\def\agamma{{B_\gamma}}

\title{ Subleading corrections to parity violating pion photoproduction }
\author{
Shi-Lin Zhu$^{a,b}$, S. Puglia$^{a,b}$,
B.R. Holstein$^{c}$, and M. J. Ramsey-Musolf$^{a,b,d}$\\
$^a$ Kellogg Radiation Laboratory 106-38, California Institute
of Technology, Pasadena, CA 91125\\
$^b$ Department of Physics, University of Connecticut,
Storrs, CT 06269 \\
$^c$ Department of Physics, University of Massachusetts,
Amherst, MA 01003 \\
$^d$ Theory Group, Thomas Jefferson National Accelerator Facility, Newport
News, VA 23606 }
\maketitle

\begin{abstract}

We compute the photon asymmetry $B_\gamma$ for near threshold parity
violating (PV) pion
photoproduction through sub-leading order. We show that sub-leading
contributions
involve a new combination of PV couplings not included in previous analyses
of hadronic PV.
We argue that existing constraints on the leading order contribution to
$B_\gamma$ --
obtained from the PV $\gamma$-decay of $^{18}$F -- suggest that the impact
of the subleading
contributions may be more significant than expected from naturalness arguments.

\medskip
PACS Indices: 11.30.Er, 11.30.Rd, 13.60.Le.
\end{abstract}

\vspace{0.3cm}
]
\pagenumbering{arabic}

\section{Introduction}

The parity violating (PV) $\pi NN$ Yukawa coupling constant
$h_\pi^{\mbox{eff}}$
is a key ingredient to the understanding of the PV nuclear
interaction\cite{ddh,haxton,fcdh,kaplan,hh95} (historically, this constant
has been denoted as $f_\pi$ in the literature). A number of hadronic
PV experiments have sought to determine the value of $h_\pi^{\mbox{eff}}$
\cite{haxton,hh95,Oers,F18,Cs123,nist}. A particularly significant result
has been obtained from measurements of photon polarization $P_\gamma$ in the PV
$\gamma$-decay of $^{18}$F:
\begin{equation}
\label{eq:flourine}
h_\pi^{\mbox{eff}}=(0.73\pm 2.3)g_\pi\ \ \ ,
\end{equation}
where $g_\pi=3.8\times 10^{-8}$ gives the scale of $g_\pi$ in the absence
of weak neutral currents\cite{ddh}. An explicit SU(6)/quark model
analysis\cite{ddh},
as well as \lq\lq naturalness" arguments (see below), would suggest
that $h_\pi^{\mbox{eff}}$ should be closer to $10 g_\pi$. The results of
the $^{18}$F
measurement, which has been repeated by five different groups, is therefore
suprising.
The nature of the $h_\pi^{\mbox{eff}}$ puzzle is further complicated by two
additional
observations:

\begin{itemize}

\item The governing PV mixing matrix element in $^{18}$F can be related by
isospin
symmetry to two body component of 
the experimental rate for the analog $\beta$ decay $^{18}$Ne$\to
^{18}$F$+e^+ + \nu_e$ \cite{haxton2,lowry}. Since $P_\gamma(^{18}$F) is
dominated by its
sensitivity to $h_\pi^{\mbox{eff}}$, the bounds in Eq. (\ref{eq:flourine})
appear to be
robust from the standpoint of many-body nuclear theory\cite{haxton}.

\item A measurement by the Boulder group of the nuclear spin-dependent PV
effects in 6S-7S 
transitions in the $^{133}$Cs atom has been used in order to extract a value for the
cesium nuclear anapole moment (AM)\cite{Cs123}. Recently, a full two-body
calculation
of the cesium AM has been used to extract constraints on the long- and
short-range
components of the PV NN interaction \cite{cesiumam}. When combined with the
constraints
on the short-range PV NN interaction, the cesium results imply a central
value for
$h_\pi^{\mbox{eff}}$ of $\sim 10 g_\pi$, in agreement with the
``naturalness'' estimate.

\end{itemize}

The status of $h_\pi^{\mbox{eff}}$ may be clarified by a slate of new
experiments --
suggested, planned, or currently underway:  ${\vec n} p\rightarrow d\gamma $
at LANSCE \cite{lan}, $\gamma^*,\gamma d\rightarrow np$ at Jefferson Lab \cite{jlab},
the rotation of polarized neutrons in helium at NIST \cite{nist} as
well as polarized Compton scattering processes \cite{bs,cck}. Since these
processes
involve one- and few-body systems, one anticipates new constraints on the PV NN
interaction
free from many-body uncertainties related to complex nuclei such as
cesium or fluorine.

If the new experiments were to confirm the present $^{18}$F constraints on
$h_\pi^{\mbox{eff}}$,
then one should attempt to understand the nucleon structure dynamics
responsible for the
reduction from its \lq\lq natural" size. At the same time, it would become
necessary to
account for the sub-leading chiral structure of the PV $\pi NN$ Yukawa
interaction and its
related observables. To that end, we recently computed the subleading
chiral contributions
to $h_\pi^{\mbox{eff}}$\cite{zhu1}. At leading order, $h_\pi^{\mbox{eff}}$
is identical to the
low-energy constant (LEC) $h_\pi^1$ appearing in the PV pion-nucleon chiral
Lagrangian\cite{kaplan}. The subleading contributions, which vanish in the
chiral limit,
involve a host of new LEC's whose effect on $h_\pi^{\mbox{eff}}$ is
fortuitously enhanced. A
similar set of LEC's appear in anapole moment contributions to the
radiative corrections to
backward angle PV $ep$ scattering. These corrections, which have recently
been determined by
the SAMPLE collaboration \cite{sample}, appear to be considerably
larger than one's
theoretical expectation\cite{zhu}. Thus, there appear to be several hints that the
chiral expansion for
hadronic PV may not behave as one na\" ively expects.

With this situation in mind, we consider in this note the subleading chiral
contributions to another
PV observable: the polarization asymmetry $B_\gamma$ for the charged pion
photoproduction
process
\begin{equation}
\overrightarrow{\gamma }\left( q^{\mu };\epsilon ^{\mu }\right)
+p(P_{i}^{\mu })\rightarrow \pi ^{+}(k^{\mu })+n(P_{f}^{\mu })\ .
\end{equation}
which will be the focus of the proposed JLab study.
Here, $q^{\mu }=\left( \omega ,{\bf q}\right) $, $P_{i}^{\mu }$, $k^{\mu
}=\left( \omega _{\pi },{\bf k}\right) $, and $P_{f}^{\mu }$ are the
center-of-mass four-momenta of photon, proton, pion and neutron,
respectively, and $\epsilon ^{\mu }$ is the photon polarization vector.
The asymmetry $B_\gamma$, which arises from the interference of PV and
parity conserving
(PC) amplitudes, was
first studied in the context of the conventional meson-exchange framework
for hadronic PV in Refs. \cite{Woloshyn,LHH}. 
Recently, Chen and Ji (CJ) proposed a measurement of $B_\gamma$ 
at the Jefferson Lab and 
recast the earlier analyses in the context
of heavy baryon chiral perturbation theory (HBCPT) \cite{j1,ijmpe}.
The authors emphasized that PV $\pi$ photoproduction accesses the PV
$NN \pi$ interaction directly, whereas in nuclear observables it
is contained within the PV NN potential.
For the threshold
region, where all external momenta are well below the chiral symmetry
breaking scale
$\Lambda_\chi = 4 \pi F_{\pi }\sim$ 1 GeV, CJ obtain the \lq\lq low-energy
theorem" for the
asymmetry:
\begin{equation}\label{old}
B_{\gamma }\left( \omega _{{\rm th}},\theta \right) =\frac{\sqrt{2}F_{\pi
}(\mu _{p}-\mu _{n})}{g_{A}m_{N}}h_{\pi}^{1}\ .
\end{equation}
and the corrections from terms higher order in the chiral expansion
were estimated to be around $20\%$ \cite{ji1}. The expression in Eq.
(\ref{old})
is consistent with the $h_\pi^{\mbox{eff}}$ dominance of $B_\gamma$ found
in Ref. \cite{LHH}.
CJ also explored the kinematic behavior of $B_\gamma$, indicating that it could be
large enough to be observed in a polarized photon beam experiment at
Jefferson Laboratory.

In this paper, we show that inclusion of subleading contributions to the PV
photoproduction amplitude leads to a chirally corrected low-energy theorem:
\begin{eqnarray}
\label{eq:newtheorem}\nonumber
B_{\gamma }\left( \omega _{{\rm th}},\theta \right) &=&\frac{\sqrt{2}F_{\pi
}}{g_{A}m_{N}}\left[\mu_p-\mu_n(1+\frac{m_\pi}{m_N})\right]h_{\pi}^{1}\\
&&-\frac{4\sqrt{2}m_\pi}{g_{A}
\Lambda_\chi}{\bar C}\ \ \ ,
\end{eqnarray}
where the $m_\pi/m_N$ represents the first recoil corrections to the
leading order
PV and PC photoproduction amplitudes and ${\bar C}$ is a new PV LEC defined
below.
In terms of chiral counting, the result
of CJ appears at ${\cal O}(p^0)$ while the corrections arising in Eq.
(\ref{eq:newtheorem}) occur at ${\cal O}(p)$. We note that the recoil
and ${\bar C}$ terms shown explicitly in Eq. (\ref{eq:newtheorem}) constitute
the complete set of subleading contributions to the PV photoproduction
amplitude, since the
effects of loops as well as pole diagrams involving decuplet intermediate
states arise at
${\cal O}(p^2)$ and beyond.

At face value, the expression in Eq. (\ref{eq:newtheorem}) indicates that
$B_\gamma$
is governed by two, rather than one, PV LEC's---$h_\pi^1$ and $\bar
C$, with associated kinematic factors of nearly equal magnitude.  The
actual
situation, however, is more subtle. The naturalness arguments which imply
$h_\pi^1$
should be $\sim 10g_\pi$ also lead one to expect ${\bar C}\sim g_\pi$.
Thus, if these
two LEC's were to have their natural size, the subleading contributions to
$B_\gamma$
would generate the anticipated 10\% effect\footnote{We thank the authors of CJ as
well as
J.L. Friar for clarification of this point.}. The results of the
$^{18}$F experiment, on the other hand, imply that $h_\pi^1$ is strongly
suppressed from
its natural scale. In this case, one would expect $h_\pi^1$ and $\bar C$ to
be of comparable
importance.  Given the present lack of a first principle QCD calculation
of these two
LEC's, it is up to experiment to settle the question. As noted in CJ, if
$h_\pi^1$ were
to have its natural size, then a 20\% determination of $B_\gamma$ may be
feasible
at Jefferson Lab. On the other hand, a null result at this precision would
be consistent
with the $^{18}$F experiment and would imply the need of additional
measurements to
separate $h_\pi^1$ and ${\bar C}$.

In the remainder of this paper, we discuss the calculations leading to our
conclusions. In section 2, we summarize the formalism for treating hadronic PV
in HBCPT. Section 3 gives the calculation of the subleading contributions
to the
PV photoproduction amplitude. In section 4, we discuss a field
redefinition, first
suggested in Ref. \cite{ji3}, which expresses the results of section 3 in a
compact
manner. In section 5, we consider the expected magnitudes of the PV LEC's,
relate
these estimates to the earlier work of Ref. \cite{LHH}, and summarize our
conclusions.

\section{Hadronic parity violation in chiral perturbation theory}

Before considering the heavy baryon expansion, it is useful to review the
relevant PC and PV Lagrangians in the fully relativistic theory. For
simplicity,
we consider only $\pi$, $N$, and $\gamma$ interactions. In this case, for PC
interactions one has
\begin{eqnarray}
\label{lpc}\nonumber
{\cal L}^{PC}&=&{1\over 4}F_\pi^2 Tr D^\mu \Sigma D_\mu \Sigma^\dag +
\bar N (i {\cal D}_\mu \gamma^\mu -m_N) N\\ \nonumber
&& + g_A \bar N A_\mu \gamma^\mu\gamma_5 N+
{e\over\Lambda_\chi}{\bar N}(c_s \\
&&+c_v\tau_3)\sigma^{\mu\nu}
N F_{\mu\nu}^++\cdots
\ \ \ ,
\end{eqnarray}
where ${\cal D}_\mu$ is the chiral and electromagnetic (EM) covariant
derivative,
$\Sigma=\exp(i\vec\tau\cdot\vec\pi/F_\pi)$, $N$ is the nucleon isodoublet
field and
\begin{eqnarray}
A_\mu&=&{i\over 2}(\xi^\dagger\partial_\mu\xi-\xi\partial_\mu\xi^\dagger)\\
F^\pm_{\mu\nu} & = &
\frac{1}{2}F_{\mu\nu}(\xi\Lambda_p\xi^{\dag}\pm\xi^{\dag}\Lambda_p
\xi) \\
\Lambda_p & = & \frac{1}{2}(1+\tau_3) \ \ \ .
\end{eqnarray}
The relevant PV Lagrangians are \cite{kaplan,zhu}
\begin{eqnarray}\label{n1} \nonumber
{\cal L}^{PV} & = & h^0_V \bar N A_\mu \gamma^\mu N
+{h^1_V\over 2} \bar N  \gamma^\mu N  Tr (A_\mu X_+^3) \\ \nonumber
&&-{h^1_A\over 2} \bar N  \gamma^\mu \gamma_5N  Tr (A_\mu X_-^3)
-{h^1_{\pi}\over 2\sqrt{2}}F_\pi \bar N X_-^3 N\\  \nonumber
&& +h^2_V {\cal I}^{ab} \bar N
[X_R^a A_\mu X_R^b +X_L^a A_\mu X_L^b]\gamma^\mu N \\
\nonumber
&&  -{h^2_A\over 2} {\cal I}^{ab} \bar N
[X_R^a A_\mu X_R^b -X_L^a A_\mu X_L^b]\gamma^\mu\gamma_5 N \\ \nonumber
&& +{c_1\over \Lambda_{\chi}} \bar N
\sigma^{\mu\nu} [F^+_{\mu\nu},
X_-^3]_+ N  +{c_2\over \Lambda_{\chi}} \bar N \sigma^{\mu\nu} F^-_{\mu\nu} N\\
&&
+{c_3\over \Lambda_{\chi}} \bar N \sigma^{\mu\nu} [F^-_{\mu\nu}, X_+^3]_+ N
\; ,
\end{eqnarray}
where
\begin{eqnarray}
\label{eq:xlr}
X_L^a & = & \xi^{\dag}\tau^a\xi \\
X_R^a & = & \xi\tau^a\xi^{\dag} \\
X_{\pm}^a & = & X_L^a\pm X_R^a\ \ \
\end{eqnarray}
and where we follow the sign convention of Ref. \cite{zhu,zhu1}.
The corresponding PC and PV Lagrangians involving $\Delta$ fields are given
in Ref.
\cite{zhu}.

Of the PV LEC's appearing in Eqs. (\ref{n1}), $h_\pi^1$ is the most familiar
and has received the most extensive theoretical
scrutiny\cite{ddh,haxton,fcdh,kaplan,hh95}.
In the context of chiral perturbation theory, the radiative corrections
to $h_\pi^1$ were discussed extensively in \cite{zhu1}, where it was
pointed out that
what nuclear PV experiments measure is an effective coupling
$h_\pi^{\mbox{eff}}$
\cite{zhu1},
which is a linear combination of LECs $h_\pi^1, h_\Delta, h_A^{(i)}$ etc.
The commonly used \lq\lq best value"---$|h_\pi^1|=5\times 10^{-7}$---quoted in
\cite{ddh}  corresponds to a large extent to a simple tree-level estimate
without loop corrections. Estimates for $h_V^i$ and $h_A^i$ have been
discussed in Refs. \cite{kaplan,zhu1}, though no analysis similar to that of
\cite{ddh} has been performed. To date, there have appeared no estimates
of the PV $NN\pi\gamma$ constants $c_i$. Nevertheless, one expects the
magnitude
of these LEC's to be roughly a few times $g_\pi$.

For purposes of computing $B_\gamma$, it is necessary to expand the non-linear
Lagrangians of Eqs. (\ref{lpc},\ref{n1}) through one $\pi$ and one $\gamma$
order. The
results for the PC interactions are familiar and we do not list them here.
For the PV Lagrangians, we also include the leading ($2\pi$) terms
proportional
to $h_A^i$---
\begin{eqnarray}
\label{eq:lpvexpand}\nonumber
{\cal L}^{PV}& = & -ih_{\pi}^{1}\pi ^{+}p^{\dagger }n \\
\nonumber
&&-{h_V\over \sqrt{2}F_\pi} \bar p\gamma^\mu n D_\mu\pi^+\\
\nonumber
&&+i\frac{h_{A}^{(1)}+\overline{h}_{A}^{(2)}}{F_{\pi }^{2}}\overline{p}%
\gamma ^{\mu }\gamma _{5}p\pi ^{+}D_{\mu }\pi ^{-}  \nonumber \\
&&+i\frac{h_{A}^{(1)}-\overline{h}_{A}^{(2)}}{F_{\pi }^{2}}\overline{n}%
\gamma ^{\mu }\gamma _{5}n\pi ^{+}D_{\mu }\pi ^{-}  \nonumber \\
&& -i e {C\over \Lambda_\chi F_\pi}\bar p
\sigma^{\mu \nu}F_{\mu\nu}n\pi^++{\mbox{h.c.}}\ \ \ ,
\end{eqnarray}
where
\begin{eqnarray}\nonumber
&h_V  =  h_V^0+\frac{4}{3}h_V^2\\
&C  = -2\sqrt{2} c_1 +{1\over \sqrt{2}}c_2\ \ \ .
\end{eqnarray}
Note that the LEC $h_V^1$  does not contribute to ${\cal L}^{PV}$ at this order.
As noted in Ref. \cite{ji3} and discussed in detail below, the effects of
the  $h_V^i$ Lagrangians on processes involving up to two pions and one
photon can be be absorbed into effective
$C$- and $h_A^i$ type Lagrangians through $2\pi$ order via an
appropriate nucleon field
redefinition. The reason is that when one integrates by parts the action
corresponding to
the $h_V$-term in Eq. (\ref{eq:lpvexpand}), the integrand vanishes by the
nucleon
equations of motion. At $3\pi$ order and beyond, however, the effects
of the
$h_V^i$ terms in Eq. (\ref{n1}) cannot be absorbed into other effective
interactions
via field redefinition. Thus, in the context of the complete non-linear PV
Lagrangian,
the $h_V^i$ remain distinct LEC's. Consequently, we keep the
$h_V$-dependence explicit
in what follows.

\section{The subleading correction to the asymmetry}

In order to maintain proper chiral counting, we use the heavy baryon
expansion of
Eqs. (\ref{lpc},\ref{n1}).
The motivation behind the use of
heavy baryon chiral perturbation theory (HBCPT) is explained in detail in
\cite{ji1}, and we follow the notations of this reference.
Since we work in the near-threshold region, we use the so-called
``small-scale'' expansion \cite{hhk}, {\it i.e.}, we
treat $\omega, \omega_\pi, |k|, m_\pi, \delta
=m_\Delta- m_N$, etc. as small quantities and characterize amplitudes
by the number of powers of these terms, {\it e.g.}, we count the
term $\omega_\pi / q\cdot k$
as being ${\cal O}(p^{-1})$. The
photon asymmetry arises from the interference of the parity conserving (PC)
and
PV  amplitudes. In Ref. \cite{ji1} the asymmetry was truncated at
leading order, {\it i.e.}, ${\cal O} (p^{0})$. In the present work we
include the ${\cal O} (p)$ correction, which arises dominantly from
the PV vector $\pi NN$ couplings. As we  show below, chiral
loops contribute to the asymmetry only at ${\cal O} (p^{2})$ and higher.
Hence, our
truncation of the chiral expansion of the asymmetry is consistent and
complete up to terms of ${\cal O} (p)$.

The PC amplitudes which describe the charged photoproduction reaction
are defined via
\begin{eqnarray}
T^{PC} &=&N^{\dagger }\left[ i{\cal A}_{1}\,{\vec \sigma }\cdot {\hat \epsilon
}+i{\cal A}_{2}{\vec \sigma }\cdot \widehat{{\bf q}}{\bf \ }\,{\hat \epsilon }
\cdot\widehat{{\bf k}}\right.   \nonumber \\
&&\left. +i{\cal A}_{3}{\vec \sigma }\cdot \widehat{{\bf k}}{\bf \ }\,
{\hat \epsilon }\cdot \widehat{{\bf k}}
 +{\cal A}_{4}{\hat \epsilon }\cdot \,\widehat{{\bf q}}\times
\widehat{{\bf k}}\right] N\ ,  \label{tpc}
\end{eqnarray}
where $N$ is the proton Pauli spinor, ${\vec \sigma }$ are the Pauli
spin matrices, and $\widehat{{\bf q}}$ and $\widehat{{\bf k}}$
are the unit vectors in the photon and pion directions respectively.
At leading order in HBCPT, we have
${\cal A}_{1}=eg_{A}/\sqrt{2}F_{\pi }$, ${\cal A}_{2}={\cal A}_{1}\omega |%
{\bf k}|/q\cdot k$, ${\cal A}_{3}=-{\cal A}_{1}{\bf k}^{2}/q\cdot k$, and
${\cal A}_{4}=0$ \cite{bkm,fearing}. As explained in \cite{ji1}
one also
requires the non-vanishing subleading order result for ${\cal A}_4$.
\begin{eqnarray}
{\cal A}_{4} &=&\frac{eg_{A}\left| {\bf k}\right| }{2\sqrt{2}F_{\pi }m_{N}}
\left[ \mu _{p}-\left( \frac{\omega }{\omega _{\pi }}\right) \mu _{n}\right]
\nonumber \\
&&-\frac{2eg_{\pi N\Delta }G_{1}\left| {\bf k}\right| }{9\sqrt{2}F_{\pi
}m_{N}}\left( \frac{\omega }{\omega -\delta }+\frac{\omega }{\omega _{\pi
}+\delta }\right) \ ,
\end{eqnarray}
where the $\Delta(1232)$ contribution has been included
explicitly. Here $G_1$
is the M1 transition moment connecting the nucleon and delta, and
$g_{\pi N\Delta}$ is the $\pi$-$N$-$\Delta$ coupling\cite{hhk}. Note that
${\cal A}_{1-3}$ is ${\cal O}(p^0)$ while ${\cal A}_4$ is ${\cal O}(p)$.

To ${\cal O} (p)$ in the chiral expansion, the PV $\gamma
p\rightarrow \pi^{+}n$ T-matrix can be written as
\begin{equation}
T^{PV}=N^{\dagger }\left[ {\cal F}_{1}\,\widehat{{\bf k}}\cdot {\hat
\epsilon }
+i{\cal F}_{2}{\vec \sigma }\cdot {\hat \epsilon }\times \,\widehat{%
{\bf q}}+i{\cal F}_{3}{\vec \sigma }\cdot {\hat \epsilon }\times \,\widehat{%
{\bf k}}\right] N\ .
\end{equation}
We then have the asymmetry
\begin{equation}\label{asy}
B_\gamma \sim \{{\cal A}_1 {\cal F}_2 +{\sin^2\theta \over 2}
[{\cal A}_3 {\cal F}_2- {\cal }A_4 {\cal F}_1 -{\cal A}_2 {\cal F}_3]
+\cos \theta {\cal A}_1 {\cal F}_3 \}\ \ \ ,
\end{equation}
where $\theta =\cos^{-1}\hat{\bf q}\cdot\hat{\bf k}$.
(Note that the nominally leading piece from the interference term ${\cal A}_{1-3}
{\cal F}_1$ vanishes if the proton target is unpolarized.)

The leading, nonvanishing
contributions to $B_\gamma$, which occur at ${\cal O}(p^0)$, are generated
by the
${\cal O}(p^0)$ terms in ${\cal A}_{1-3}$ interfering with the ${\cal
O}(p^0)$ terms
in ${\cal F}_2$, and by the ${\cal O}(p)$ term in ${\cal A}_4$ interfering
with
the ${\cal O}(p^{-1})$ term in ${\cal F}_1$. The leading order PV
contributions
to ${\cal F}_{1,2}$ arise from the insertion of the PV Yukawa $\pi NN$
vertex of Eq. (\ref{eq:lpvexpand}) in FIG. 1 (a), (c), (d).
The results, given in Ref. \cite{ji1}, are
\begin{equation}
\label{eq:leadingPV}
{\cal F}_{1}=-\frac{eh_{\pi }^{1}\left| {\bf k}\right| }{q{\bf \cdot }k}%
\ ,\ {\cal F}_{2}=-\frac{eh_{\pi }^{1}}{2m_{N}}\left[ \mu _{p}-\left(
\frac{\omega }{\omega _{\pi }}\right) \mu _{n}\right] \ .
\end{equation}
where ${\cal F}_1, {\cal F}_2$ are ${\cal O}(p^{-1}), {\cal O}(p^0)$
respectively.

Subleading contributions to $B_\gamma$ are generated by ${\cal O}(p)$ and
${\cal O}(p^2)$ terms in ${\cal A}_{1-3}$ and ${\cal A}_4$, respectively,
intefering with the amplitudes in Eq. (\ref{eq:leadingPV}), and by
${\cal O}(p)$
contributions in ${\cal F}_{2,3}$ interfering with the ${\cal O}(p^0)$
terms in
${\cal A}_{1-3}$. The subleading PC contributions have been computed in
\cite{fearing}. We refer to the detailed expressions for these corrections in
that work, which we employ in our numerical analysis below. Of greater
interest
are the ${\cal O}(p)$ PV amplitudes involving new LEC's. These
contributions, which are generated by the $h_V$ and $C$ terms in eq.
(\ref{eq:lpvexpand}),
contribute to both the pole
diagrams FIG. 1 (c), (d) and the seagull diagram FIG. 1 (b).
We have
\begin{eqnarray}\label{pole}\nonumber
&{\cal F}_{1}={\cal F}_3=0\; , \\
&{\cal F}_{2}=\frac{eh_V}{2m_{N}}{\omega_\pi\over
\sqrt{2}F_\pi}\left[ \mu_{p}- {\omega\over \omega_\pi}
  \mu _{n}-{\omega\over \omega_\pi}\right]
 +\frac{2eC} {\Lambda_\chi}\frac{\omega}{F_\pi}\; .
\end{eqnarray}
The contribution from FIG. 1 (b) cancels exactly those from FIG. 1 (c), (d)
where the $\gamma NN$ vertex is minimum coupling \footnote{We thank J.-W. Chen
and X. Ji for pointing out this cancellation to us.}.

According to the expression in Eq. (\ref{pole}), the $h_V$ and $C$
contributions
to ${\cal F}_2$ carry distinct kinematic dependences, a feature which might
suggest
using the $\omega$-dependence of $B_\gamma$ to separate the two LEC's. Such
a program
would be misguided, however. As we show below, the kinematic behavior
generated by
the $h_V$ and $C$ interactions is identical when a fully relativistic
framework is
used to compute the PV amplitudes. The result in this case is
\begin{equation}
\label{polerel}
{\cal F}_2 =  {2e{\bar C}\over\Lambda_\chi} {\omega\over F_\pi}
\end{equation}
with
\begin{equation}
{\bar C}  =  C + {\Lambda_\chi\over m_N}
\left({\kappa_p-\kappa_n\over 4\sqrt{2}}\right) h_V \ \ \ .
\end{equation}
Here $\kappa_i$ are the anomalous nuclon magnetic moments, as
distinguished from the full moments $\mu_i$ used to this point.
The apparent difference between  Eqs. (\ref{pole},\ref{polerel})
is a artifact of truncating the $1/m_N$ expansion at this order in
HBCPT---to this order of the chiral expansion the photon and pion
energies are equal.
In what
follows, then, we adopt the result in Eqs. (\ref{polerel}).

In addition to the ${\cal O}(p)$ contributions from $h_V$ and $C$, ${\cal
F}_2$
receives an ${\cal O}(p)$ contribution involving $h_\pi^1$ generated by the
$1/m_N$
corrections to the nucleon propagator and $\gamma NN$ vertex in the
pole
amplitudes. We include these corrections in the asymmtry formulae
below. Other possible  contributions to
the PV amplitudes include PV $\gamma N
\Delta$ and $\pi N \Delta$ interactions. However, both contribute at ${\cal
O}(p^2)$,
which is higher than the order at which we are truncating. Similarly
 chiral loop contributions to ${\cal A}_{1-4}, {\cal F}_1, {\cal F}_2,
{\cal F}_3$ appear at ${\cal O}(p^2)$, ${\cal O}(p)$, ${\cal O}(p^2)$
or higher, respectively. Consequently, chiral loops do not contribute to the
asymmetry until at least ${\cal O}(p^2)$. All such contributions are
then higher order and
can be dropped.

The resultant photon asymmetry at order ${\cal O}(p)$ reads then
\begin{eqnarray}
\nonumber
&B_\gamma(\omega, \theta)  =  {\sqrt{2} h_\pi^1 F_\pi\over g_A m_N {\cal G}}
\Biggl\{\Biggl[\left(1-{\omega\over 2m_N}\right)\mu_p \\ \nonumber
& \ \ \ -\left({\omega\over\omega_\pi}\right)\left(1-{|{\bf k}|^2\over 2
m_N\omega_\pi}
\right)\mu_n\Biggr]\left( 1-\sin^2\theta\frac{ {\bf k}^{2}}{q
{\bf \cdot }k}\right) \\ \nonumber
&+\frac{2}{9}g_{\pi N\Delta }G_{1}\sin ^{2}\theta \frac{{\bf k}^{2}}{
g_{A}q{\bf \cdot }k}\left( \frac{\omega }{\omega -\delta }+\frac{\omega }{
\omega _{\pi }+\delta }\right)\\ \nonumber
&+2\left({\omega\over\omega_\pi}\right){|{\bf k}|\over m_N}\mu_n
\left( \cos\theta -\sin ^{2}\theta \frac{\omega |{\bf k}|}{2q {\bf \cdot
}k}\right)\Biggr\}\\
&
-{4\sqrt{2}{\bar C}\over g_A\Lambda_\chi{\cal G}}\left( 1-\sin
^{2}\theta \frac{{\bf
k}^{2}}{2q {\bf \cdot }k}\right)+\cdots\ \ \ .\label{eq:asy}
\end{eqnarray}
where the ellipses indicate the PC $1/m_N$ contributions of
$T^{PC}$ in Ref. \cite{fearing} and 
\begin{equation}
{\cal G}=1-\sin^{2}\theta \frac{{\bf k}^{2}}{q {\bf \cdot }k}
[1-{({\bf q}-{\bf k})^2\over 2q\cdot k}]
\; .
\end{equation}
At threshold---$|{\bf k}|=0$---Eq. (\ref{eq:asy}) becomes 
the low energy theorem for the photon
asymmetry given in Eq. (\ref{eq:newtheorem}).

\section{Field redefinition and physical observables}

In response to an earlier version of this paper, CJ observed that one may
obtain the subleading PV contributions to $B_\gamma$ involving $\bar C$
entirely from the diagram (b) in Fig. 1 after a suitable redefinition of the
nucleon fields \cite{ji3}. This simplification arises because the $h_V$-terms
in Eq. (\ref{eq:lpvexpand}) vanish for on-shell nucleons after integration
by parts. As discussed in Ref. \cite{field}, the effects of interactions which
vanish by the equations of motion can always be absorbed into contact
interactions via field redefinition. In the present case, the redefinition
proposed by CJ is
\begin{eqnarray}\label{red}\nonumber
&p = {\tilde p} - {i\over \sqrt{2} F_\pi}h_V \pi^+ {\tilde n} \, ,\\
&n =  {\tilde n} - {i\over \sqrt{2} F_\pi}h_V \pi^- {\tilde p} \, .
\end{eqnarray}
The resultant PV Lagrangian ${\tilde{\cal L}}^{PV}$ is
\begin{eqnarray}\label{20}\nonumber
{\tilde{\cal L}}^{PV} &=&-ih_{\pi }^{1}\pi ^{+}\overline{\tilde p}{\tilde
n}  \\
&&+i\frac{h_{A}^{1}+\overline{h}_{A}^{2}}{F_{\pi }^{2}}\overline{\tilde
p}\gamma ^{\mu
}\gamma _{5}{\tilde p}\pi ^{+}D_{\mu }\pi ^{-}  \nonumber \\
&&+i\frac{h_{A}^{1}-\overline{h}_{A}^{2}}{F_{\pi }^{2}}\overline{\tilde
n}\gamma ^{\mu
}\gamma _{5}{\tilde n}\pi ^{+}D_{\mu }\pi ^{-}  \nonumber \\
&&-ie\frac{\overline{C}}{\Lambda_\chi F_{\pi }}\overline{\tilde p}\sigma
^{\mu \nu
}F_{\mu \nu }
{\tilde n}\pi
^{+}+{\rm h.c.}+\cdots \ .
\end{eqnarray}
where
\begin{equation}\label{21}
\overline{h}_{A}^{(2)}=h_{A}^{(2)}-\frac{g_{A}}{2}h_{V}\ ,\quad \overline{C}%
=C+{\Lambda_\chi\over m_N}\left(
\frac{\kappa _{p}-\kappa _{n}}{4\sqrt{2}}\right) h_{V}\ .
\end{equation}
Note that in ${\tilde{\cal L}}^{PV}$, the $h_V$-terms have been eliminated, and
their effect absorbed into the LEC ${\bar C}$ and $\overline{h}_{A}^{(2)}$
introduced earlier\footnote{Our
relative phase between $C$ and $h_V$ in $\bar C$ differs from Ref.
\cite{ji3}.}.
In terms of physical observables involving up to two $\pi$ and one $\gamma$, it
is not possible to determine $h_V$ from $C$. In particular, as noted in Ref.
\cite{zhu}, the PV NN potential contains no dependence on $h_V$.

The question remains as to whether the $h_V^i$ constitute distinct LEC's in
the context of the full nonlinear Lagrangian of Eq. (\ref{n1}), or whether
their
effects can be entirely absorbed into other LEC's. In the following, we
address
this question using the simplest unitarized version of the transformation in
Eq. (\ref{red}). We show that at $3\pi$ order, it is not possible to
eliminate
the $h_V^i$ effects in terms of other LEC's. We give a general proof
of this
result in the Appendix. In principle, then, one could use
an appropriate PV $3\pi$ process ({\em e.g.}, the analyzing power for
$\pi^-{\vec p}\to \pi^+\pi^- n$) to separate the $h_V^i$ and $C$. In
practice,
measurements of multi-pion processes would be extremely difficult at best.

To illustrate this result, consider the unitary transformation
\begin{equation}\label{unit}
N= V_1 {\tilde N}
\end{equation}
to eliminate the leading linear term after expansion of PV
vector pieces in (\ref{n1}).
The explicit expression of $V_1$ is
\begin{eqnarray}
\label{unit2}
&V_1=e^{-{i\over F_\pi} {\hat O_1}}=V e^{-{i h_V^1\over F_\pi}\pi^0} \\
&V=e^{-{i\over F_\pi} {\hat O}} \\
& {\hat O_1} ={\hat O} + h_V^1\pi^0 {\hat 1} \\
& {\hat O} ={h_V^0 \over 2} \pi^i \tau^i + {4\over 3} h_V^2
 \left( \begin{array}{ll} 2\pi^0 & {\pi^+\over \sqrt{2}}\\
 {\pi^-\over \sqrt{2}} & -2\pi^0 \end{array} \right)
\end{eqnarray}
The difference between the field redefinition Eq. (\ref{red}) and
Eq. (\ref{unit}) is two-fold. The latter is unitary and also takes into
account
the PV vector $\pi^0 NN$ interaction.

It is useful to collect some relevant terms of the redefined Lagrangians
containing the
nucleon field ${\tilde N}$. For the  strong and
electromagnetic part we have
\begin{eqnarray}
\label{lpcred}
\nonumber
& {\tilde{\cal L}}^{PC}= \bar{\tilde N} (i D_\mu \gamma^\mu -m_N) {\tilde N}
+ \bar{\tilde N} [V_1^\dag i D_\mu V_1] \gamma^\mu  {\tilde N} \\ \nonumber
&+ \bar{\tilde N} [V^\dag i V_\mu V] \gamma^\mu  {\tilde N}
+ g_A \bar{\tilde N} [V^\dag A_\mu V ]\gamma^\mu \gamma_5 {\tilde N} \\
&+{e\over \Lambda_{\chi}}\bar {\tilde N}[ V^\dag (c_s +c_v\tau_3)
\sigma^{\mu\nu}
F^+_{\mu\nu} V ]{\tilde N} +\cdots
\end{eqnarray}
where $V_\mu$ is the chiral connection.

For the originally weak interaction  we have
\begin{eqnarray}\label{lpvred} \nonumber
&{\tilde{\cal L}}^{PV}=h^0_V \bar {\tilde N} V^\dag A_\mu V \gamma^\mu
{\tilde N}
+{1\over 2} h_V^1\bar {\tilde N} \gamma^\mu {\tilde N}  Tr (A_\mu X_+^3) \\
\nonumber
&-{1\over 2} h_A^1\bar {\tilde N}  \gamma^\mu \gamma_5 {\tilde N}  Tr (A_\mu
X_-^3)
-{1\over 2\sqrt{2}}h_\pi^1F_\pi \bar {\tilde N}[V^\dag X_-^3 V]{\tilde
N}\\  \nonumber
&+h^2_V {\cal I}^{ab} \bar {\tilde N}
V^\dag [X_R^a A_\mu X_R^b +X_L^a A_\mu X_L^b] V \gamma^\mu {\tilde N} \\
\nonumber
& \ \ \ -{1\over 2}h_A^2 {\cal I}^{ab} \bar {\tilde N} V^\dag
[X_R^a A_\mu X_R^b -X_L^a A_\mu X_L^b] V\gamma^\mu\gamma_5 {\tilde N} \\
\nonumber
& +{1\over \Lambda_{\chi}} c_1\bar {\tilde N}
\sigma^{\mu\nu} [F^+_{\mu\nu},
X_-^3]_+ {\tilde N}  +{1\over \Lambda_{\chi}}c_2
\bar {\tilde N} \sigma^{\mu\nu} F^-_{\mu\nu} {\tilde N}\\
&
+{1\over \Lambda_{\chi}} c_3\bar {\tilde N} \sigma^{\mu\nu} [F^-_{\mu\nu},
X_+^3]_+
{\tilde N}
\; ,
\end{eqnarray}

Now expand Eqs. (\ref{lpcred}) and (\ref{lpvred}) in ${1/F_\pi}$. The leading
term arising from
\begin{equation}
\bar{\tilde N} [V_1^\dag i D_\mu V_1] \gamma^\mu  {\tilde N}
\end{equation}
in Eq. (\ref{lpcred}) entirely cancels the  $1\pi$ $h_V^i$ terms in Eq.
(\ref{lpvred}),
recovering the results of Eqs. (\ref{20}-\ref{21}).
The potential sources of $3\pi$ PV interactions include the following:

\medskip
\noindent (1) expansion of the term in Eq. (\ref{lpcred})
$\bar{\tilde N} [V_1^\dag i D_\mu V_1] \gamma^\mu  {\tilde N}$
in Eq. (\ref{lpcred}). The result is ${\cal O}(G_F^3)$.

\medskip
\noindent (2) expansion of the term in Eq. ({\ref{lpcred})
$\bar{\tilde N} [V^\dag i V_\mu V] \gamma^\mu  {\tilde N}$, which is
linear in $h_V^i$, $i=0, 2$ only [${\cal O}(G_F)$];

\medskip
\noindent (3) expansion of $A_\mu, X_{\pm}^3, X_{L,R}^a$ operators
 in Eq.({\ref{lpvred}) to third order,
which is linear in $h_V^i$, ($i=0, 1, 2$) and $h_A^i, h_\pi^1$ ($i=1,2$)
[${\cal
O}(G_F)$];

\medskip
\noindent (4) expansion of $V$ and $V^\dag$ operator in Eq.({\ref{lpvred})
to second order,
which is cubic in $h_V^i$, $i=0, 2$ only [${\cal O}(G_F^3)$].

\medskip
\noindent (5) expansion of the $\bar{\tilde N} [V^\dag A_\mu V]
\gamma^\mu\gamma^5  {\tilde
N}$ and $\bar {\tilde N}[ V^\dag (c_s +c_v\tau_3)  \sigma^{\mu\nu}
F^+_{\mu\nu} V ]{\tilde N}$ terms in to third order [${\cal O}(G_F^2)$ and
${\cal O}(G_F)$, respectively].

\medskip
\noindent (6) expansion of the $c_i$-terms in Eq. (\ref{lpvred}) to third order
[${\cal O}(G_F)$].

Prior to the applying the transformation (\ref{unit},\ref{unit2}), the only PV
$NN\pi\pi\pi$  contact interactions arise from the $h_V^i$-terms in (3).
After field
redefinition, one must add up all six contributions. Note that those
arising from (5), (6)
and the $h_\pi^1, h_A^i$-terms in (3) contain a different Lorentz structure
than the
$h_V^i$
terms in (3) and therefore cannot cancel the latter. Similarly, since the
$h_V^i$ $3\pi$ terms in (3) arise at ${\cal O}(G_F)$, they cannot be
cancelled by the contributions from (1) and (4). Thus, at ${\cal O}(G_F)$, the
only $3\pi$ contributions involving $\bar{\tilde N}\gamma_\mu{\tilde N}$
arise from (2) and the $h_V^i$ terms in (3). Note that (2) contains no terms
involving $h_V^1$. Hence, the $3\pi$ term proportional to $h_V^1$
appearing
in (3) cannot be removed by the transformation Eq. (\ref{unit}).

For the terms proportional to $h_V^0$ we obtain from (2)
\begin{equation}
-{h_V^0\over 2F_\pi^3} [\pi, [\pi, D_\mu \pi]] \; .
\end{equation}
where $\pi ={1\over 2}\pi^i \tau^i$, while (3) yields
\begin{equation}
+{h_V^0\over 6F_\pi^3} [\pi, [\pi, D_\mu \pi]] \; .
\end{equation}
Their sum is
\begin{equation}
-{h_V^0\over 3 F_\pi^3} [\pi, [\pi, D_\mu \pi]] \; .
\end{equation}
The $3\pi$ PV vector $h_V^0$ contact term does not
vanish after field redefinition. A similar result holds for $h_V^2$.

As we show in the Appendix, one may remove the  $1\pi$ $h_V^i$ terms
by a more general field redefinition than given by Eqs.
(\ref{unit},\ref{unit2}).
Nevertheless, it is still not possible to remove the $3\pi$ terms
proportional to the $h_V^i$ (the arguments of the proof are similar to
those above, but more tedious in the details). Thus, we conclude that the
$h_V^i$ constitute distinct and, in principle measurable LEC's in the nonlinear
chiral theory of Eqs. (\ref{lpc},\ref{n1}). While one could compute observables
in either formulation of the theory (with or without the field redefinition)
and obtain identical results, the structure of Lagrangian is more
cumbersome after
application of Eq. (\ref{unit}): there appear several new interaction
vertices, including small [${\cal O}(G_F^2)$] parity-conserving non-derivative
interactions; the chiral transformation properties are less transparent than
in the original version of the theory; and the fields ${\tilde N}$ annihilate
nucleon states of mixed parity. Consequently, we retain the
original form of ${\cal L}^{PV}$ given in Refs. \cite{kaplan,zhu}.

\section{Scale of the LEC's}

Given that $h_\pi^1$ and $\bar C$ appear in $B_\gamma$ with nearly equal
weight, it would be useful to have in hand a theoretical expectation for the
magnitudes of these LEC's. A simple estimate can be obtained by applying
the \lq\lq na\"\i ve dimensional analysis"
of Ref. \cite{gm}. For strong and EM interations, effective
interactions
scale with $F_\pi$ and $\Lambda_\chi$ as
\begin{equation}
\label{eq:georgi}
(\Lambda_\chi F_\pi)^2 \ \times\ \left({{\bar N} N\over\Lambda_\chi
F_\pi^2}\right)^k
\left({\pi\over F_\pi}\right)^l \left({D_\mu \over\Lambda_\chi}\right)^m \ \ \ ,
\end{equation}
where $k,l, m$ are integers and $ D_\mu$ is the covariant derivative.
 For weak interactions, the same counting applies,
multiplied by an overall scale of
\begin{equation}
\label{eq:gpi}
g_\pi\sim {G_F F_\pi^2\over 2\sqrt{2}} \ \ \ .
\end{equation}
Thus, one would expect the strength of the PV $NN\pi$ Yukawa interaction to be
given by Eqs. (\ref{eq:georgi},\ref{eq:gpi}) with $k=1$, $l=1$, $n=0$:
\begin{equation}
\label{eq:hpiexpect}
{\Lambda_\chi\over F_\pi}\ g_\pi = 4\pi g_\pi\ \ \ .
\end{equation}
Since the definition of the Yukawa interaction in Eq. (\ref{n1}) contains
no explicit
factors of $\Lambda$ or $F_\pi$, one expects the natural size of this LEC
to be given by Eq.
(\ref{eq:hpiexpect}).  Similarly, the ${\bar C}$ interaction, which involves
$k=1$, $l=1$, $m=2$, should scale as
\begin{equation}
\label{eq:cbarexpect}
{1\over \Lambda_\chi F_\pi} \ g_\pi \ \ \ .
\end{equation}
However, since the PV $NN\pi\gamma$ contact interaction in
Eq. (\ref{lpvred}) 
already contains the
explicit factors $1/\Lambda_\chi$ and $1/F_\pi$, the coefficient -- $\bar
C$ --
should be roughly of size $g_\pi$.

It is useful to compare these expectations with results of model
calculations as
well as with experiment. The benchmark SU(6)/quark model calculation of
Ref. \cite{ddh},
updated in Ref. \cite{fcdh}, gives a \lq\lq best" estimate for
$h_\pi^{\mbox{eff}}$ of
$(7-12)\times g_\pi$ -- roughly commensurate with the expection of Eq.
(\ref{eq:hpiexpect}).
That analysis, however, allows for the Yukawa coupling to be as small as
zero and as
large as $(20-30)\times g_\pi$, owing to uncertainties associated with
various SU(6)
reduced matrix elements and quark model inputs. To date, no 
estimate of $\bar
C$ has been performed. A simple estimate can be made, however, by assuming the
short-distance PV
physics is saturated by $t$-channel vector meson exchange. In the purely
mesonic sector,
one may understand the magnitudes of the ${\cal O}(p^4)$ LEC's $L_i$ using
vector meson
saturation. For the baryon sector, the same framework was used to estimate
the sub-leading
contributions to the nucleon anapole moment \cite{zhu}. In the present
instance,
an illustrative contribution in this context is given in Fig. 2, where the
$\bar
C$-amplitude is generated by the PV $\rho NN$ interaction.
For the $\rho \pi\gamma$ vertex we use the Lagrangian:
\begin{equation}
{\cal L}_{\rho \pi \gamma}^{PC}=e {g_{\rho \pi \gamma}\over 4m_\rho}
\epsilon^{\mu\nu\alpha\beta}
F_{\mu\nu}G^-_{\alpha\beta} \pi^+ +\cdots
\end{equation}
where $G_{\alpha\beta}=\partial_\alpha \rho_\beta - \partial_\beta
\rho_\alpha$.
From the $\rho$ radiative decay width \cite{pdg} we have
$ |g_{\rho \pi \gamma}|=0.6$, and
for the PV $\rho NN$ interaction we follow Ref. \cite{ddh}, writing
\begin{equation}
{\cal L}_{\rho NN}^{PV}=\sqrt{2}(h_\rho^0- {h_\rho^2\over 2\sqrt{6}})
[\bar p \gamma_\mu\gamma_5 \rho^+ n + H.c. +\cdots]
\end{equation}
Invoking VMD we have
\begin{eqnarray}\nonumber
{\bar C} & \sim & -{g_{\rho \pi \gamma}\over \sqrt{2}} {\Lambda_\chi F_\pi
m_\pi \over
m_\rho^3}(h_\rho^0- {h_\rho^2\over 2\sqrt{6}})\\
&\sim& -0.35 g_\pi \ \ \ ,
\end{eqnarray}
where we have used the DDH \lq\lq best values" $h_\rho^0=-30g_\pi,
h_\rho^2=-25g_\pi$
\cite{ddh}. Presumably, other heavy mesons contribute with comparable
strength.
In this simple vector meson saturation picture, then, the size of $\bar C$ is
consistent with the expectation in Eq. (\ref{eq:cbarexpect}). We note that the
authors of Ref. \cite{LHH} adopted similar picture for the short-distance PV
physics, treating the $\rho$ and $\omega$ as explicit dynamical degrees of
freedom.

As stated at the outset of this work, the quandry for the effective field
theory
treatment of $B_\gamma$ is that the constraints on $h_\pi^1$ from the
$P_\gamma(^{18}$F)
measurements imply that this coupling is considerably suppressed from its \lq\lq natural"
scale\footnote{The $^{18}$F result is also consistent with the combined
results of PV
asymmetry measurements with $^{19}$F, $p+\alpha$, and $pp$ processes (see,
{\em e.g.},
Ref. \cite{cesiumam}).}. While the analysis of Refs. \cite{ddh,fcdh} can
accomodate the
$^{18}$F result, one has a more difficult task of explaining this result
using effective
field theory alone, without reference to the underlying dynamics of strong
and weak
interactions. Nevertheless, taking the $^{18}$F result at face value
implies that in the
HBCPT treatment  of one- and few-body PV processes nominally sensitive to
the PV $\pi NN$
Yukawa coupling, one must also take into consideration subleading PV
contributions as we
have done for $B_\gamma$.  Disentangling the short-distance physics
responsible for these
subleading effects then remains an interesting and unsolved problem for both
theory and
experiment.

\section*{Acknowledgment}
We thank J.-W. Chen, X. Ji, and J.L. Friar for several helpful conversations.
This work was supported in part under U.S. Department of Energy contracts
\#DE-AC05-84ER40150 and \#DE-FG02-00ER41146, the National Science Foundation,
and a National Science Foundation Young Investigator Award.

\section{Appendix A}

We present here a general proof that the PV 3$\pi$ vector
interaction vertex (proportional to the $h_V^i$) cannot be removed by any
unitary
transformation $U$. To simplify notation, we absorb the
factor $1/F_\pi$ into the pion field. From now on it is
understood that
\begin{equation}
\pi = {1\over F_\pi} \pi^i {\tau^i\over 2}\ \ \ .
\end{equation}
Since the transformation is unitary, we have
\begin{equation}
{\hat U} =e^{-i{\hat F}}
\end{equation}
\begin{equation}
{\hat F} ={\hat F}^\dag
\end{equation}
The operator ${\hat F}$ can be expanded in terms of
the number of pions. Since ${\hat F}$ should not carry
explicit Lorentz indices, any derivatives should appear in pairs.
Because we are discussing $3\pi$ PV vertex with only one derivative
in the present case, the possible derivative terms
are irrelevant here. Consequently, we omit them from the following
discussion. We
also consider explicitly only the $h_V^{0,1}$ contributions; the arguments
involving
$h_V^2$ are similar, but considerably more tedious.

Expand ${\hat F}$:
\begin{equation}
{\hat F} ={\hat O}_1 + {\hat O}_2 + {\hat O}_3+ \cdots\ \ \ ,
\end{equation}
where ${\hat O}_n$ contains products of $n\pi$ fields.
The leading term ${\hat O}_1$ is needed to remove the $1\pi$ PV vector
linear term. Its structure is fixed and of ${\cal O} (G_F)$ as discussed in
Section 4. The remaining terms ${\hat O}_n$, $n>1$ could, in principle,
be of ${\cal O} (G_F^0)$. In the present case, we need to consider only
the terms through $n=3$.
The most general forms of ${\hat O}_2$, ${\hat O}_3$ read
\begin{eqnarray}\nonumber
{\hat O}_2 = (a_1 \pi^+\pi^- +a_2 \pi^0\pi^0){\hat 1} & \\ \nonumber
+(a_3 \pi^+\pi^- +a_4 \pi^0\pi^0)\tau_3 & \\ \nonumber
+ a_5 \pi^0 (\pi^+\tau_+ +\pi^- \tau_-) &\\
+ i a_6 \pi^0 (\pi^+\tau_+ -\pi^- \tau_-)
\end{eqnarray}
\begin{eqnarray}\nonumber
{\hat O}_3 = (b_1 \pi^+\pi^- +b_2 \pi^0\pi^0)\pi^0 {\hat 1} & \\ \nonumber
+(b_3 \pi^+\pi^- +b_4 \pi^0\pi^0)\pi^0 \tau_3 & \\ \nonumber
+ (b_5 \pi^+\pi^- +b_6 \pi^0 \pi^0) (\pi^+\tau_+ +\pi^- \tau_-) &\\
+ i(b_7 \pi^+\pi^- +b_8 \pi^0 \pi^0) (\pi^+\tau_+ -\pi^- \tau_-)
\end{eqnarray}
where $a_{1-6}, b_{1-8}$ are real numbers.

Now perform the unitary transformation
\begin{equation}
\label{eq:genunit}
N = {\hat U} {\tilde N}\ \ \ .
\end{equation}
The possible sources of PV vector $3\pi$ vertices in the transformed
Lagrangians  are the same
as discussed in the section 4 (items 1-6, but the order in $G_F$ is
not {\em a priori} fixed here).  In addition, we must also expand the
$X_{\pm}^a$ along with $A_\mu$ in item (3). As was done previously, we may
neglect those terms whose Lorentz structure differs from ${\bar{\tilde N}}
\gamma_\mu {\tilde N}$. Thus, we consider only the vector terms arising from
(1)-(4) (with $V\to{\hat U}$). From (1) we obtain the three $\pi$ contribution
\begin{equation}
\label{111}\nonumber
{\hat U}^{\dag} iD_\mu {\hat U} = D_\mu {\hat O}_3 +
i [{\hat O}_1, D_\mu {\hat O}_2]
-i [D_\mu {\hat O}_1,  {\hat O}_2]
+{\cal O}({\hat O}_1^3)\ \ \ ,
\end{equation}
where the ${\hat O}_1^3$ term is ${\cal O}(G_F^3)$ and may be neglected.
Since the component of ${\hat O}_1$ proportional to $h_V^1$ is independent of
the $\tau^a$, it does not contribute to the commutators in Eq. (\ref{111}).
Hence, we may replace ${\hat O}_1\to {\hat O}$ in the expression above. Since
the ${\hat O}_{2,3}$ are may be of ${\cal O}(G_F^0)$, item (1) will generate
relevant $3\pi$ terms under the general unitary transformation.

From item (2) we obtain
\begin{equation}
\label{222}
{\hat U}^\dag i V_\mu {\hat U} =-[{\hat O}_1, V_\mu^{(2)}]+\cdots=-[{\hat
O}, V_\mu^{(2)}]+
\cdots\ \ \ ,
\end{equation}
where $V_\mu^{(2)}$ denotes the  $2\pi$ terms in $V_\mu$.

Next, consider the contributions from item (3), including the expansion of the
$X_{\pm}^3$. The term proportional to
$h_V^1$ [we neglect the ${\cal O}(G_F^3)$] terms is
\begin{equation}
\label{333}
 {1\over 2} h_V^1 Tr [A_\mu X_+^3]
 = {2\over 3}h_V^1 [\pi^i \pi_i D_\mu \pi^0 -\pi^0 \pi_i D_\mu\pi^i]
\end{equation}
which does not contain $\tau_\pm, \tau_3$.
In order to remove the above term we also
need similar terms with ${\hat 1}$ structure from Eq. (\ref{111})-(\ref{222}).
The commutators
never contribute to ${\hat 1}$ structure. So the only possibility is
the isoscalar piece of ${\hat C}$,
\begin{equation}
\label{333p}
D_\mu [b_1\pi^+\pi^-\pi^0 +b_2 \pi^0\pi^0\pi^0]
\end{equation}
which is a total derivative of $3\pi$ fields, and each term is symmetric
under field permutations.
However, Eq. (\ref{333}) does not display such permutation symmetry. In
other words,
Eqs. (\ref{333}) and (\ref{333p}) cannot completely cancel each other.
Thus, the $3\pi$ $h_V^1$
piece will remain under any unitary transformation.

Now consider the $h_V^0$ term in item (3).
Expansion of the $A_\mu$
operator in ${\hat U}^\dag A_\mu {\hat U}$ in Eq. (\ref{lpvred}) leads to
\begin{eqnarray}
\label{444}
\nonumber
\sim {1\over 6} h_V^0 [\pi, [\pi, D_\mu \pi]] & \\  \nonumber
={1\over 6}h_V^0 \{ {\tau_3\over 2} [2\pi^+\pi^-D_\mu \pi^0
-\pi^0 D_\mu (\pi^+\pi^-) ]  & \\  \nonumber
+{\tau_+\over \sqrt{2}} [-\pi^+ (\pi^+D_\mu \pi^-
-\pi^-D_\mu \pi^+) & \\  \nonumber
+\pi^0 (\pi^0D_\mu \pi^+ -\pi^+D_\mu \pi^0)]& \\  \nonumber
+{\tau_-\over \sqrt{2}} [-\pi^- (\pi^-D_\mu \pi^+
-\pi^+D_\mu \pi^-) & \\
+\pi^0 (\pi^0D_\mu \pi^- -\pi^-D_\mu \pi^0)] \}\ \ \ .
\end{eqnarray}

Finally, from item (4) we obtain for the $h_V^0$ contribution
\begin{eqnarray}
\label{555}\nonumber
\sim i h_V^0 [{\hat O}_2, A_\mu] +{\cal O} ( {\hat O}_1^3)+\cdots &\\
=i h_V^0 [{\hat O}_2, A_\mu] +{\cal O} ( G_F^3)+\cdots\ \ \ .
\end{eqnarray}

Now we require the explicit three $\pi$ expressions from Eqs. (\ref{111},
\ref{222},\ref{555}) [items (1),(2), (4)] in addition to the
expression in Eq. (\ref{444}) [item (3)]. These expressions are linear in
the
$\tau_i$ and
${\hat 1}$.
For clarity, we first focus on the terms involving $\tau_3$. From Eq.
(\ref{111})
we have
\begin{eqnarray}\label{48}\nonumber
\sim D_\mu {\hat O}_3|_{\tau_3}
+i h_V^0 [ \pi, D_\mu {\hat O}_2]|_{\tau_3} &\\
\nonumber
-i h_V^0 [D_\mu \pi, {\hat O}_2]|_{\tau_3} &\\
\nonumber
= D_\mu (b_3 \pi^+\pi^-\pi^0 +b_4 \pi^0\pi^0\pi^0) \tau_3 &\\ \nonumber
+i\sqrt{2} a_5  (h_V^0+{4\over 3}h_V^2)
[\pi^+D_\mu \pi^-
-\pi^- D_\mu \pi^+]\pi^0 \tau^3 &\\
+\sqrt{2}a_6 h_V^0 \pi^+\pi^-D_\mu\pi^0\tau_3
\end{eqnarray}
where have used the following identity:
\begin{eqnarray}\nonumber
[\pi, D_\mu {\hat O}_2]|_{\tau_3}={a_5\over \sqrt{2}}
[\pi^+D_\mu \pi^- -\pi^- D_\mu \pi^+]\pi^0 \tau^3 &\\
-{ia_6\over \sqrt{2}}[2\pi^+\pi^-D_\mu\pi^0 +\pi^0 D_\mu
(\pi^+\pi^-)]\tau_3
\ \ .
\end{eqnarray}
The contribution from Eq. (\ref{222}) [item (2)] is
\begin{eqnarray}
\label{455}\nonumber
-[{\hat O}, V_\mu]|_{\tau_3}=-{1\over 4}h_V^0
[2\pi^+\pi^- D_\mu \pi^0  &\\
-\pi^0 D_\mu (\pi^+\pi^-)]\tau_3
\end{eqnarray}
while from Eq. (\ref{444}) [item (3)] we obtain
\begin{equation}\label{46}
+{1\over 12} h_V^0 [2\pi^+\pi^- D_\mu \pi^0
-\pi^0 D_\mu (\pi^+\pi^-)]\tau_3\ \ \ .
\end{equation}
Finally,  Eq. (\ref{555}) [item (4)] gives
\begin{eqnarray}\label{47}\nonumber
i h_V^0 [D_\mu \pi, {\hat O}_2]|_{\tau_3}
= -i h_V^0 {a_5\over \sqrt{2}}
[\pi^+D_\mu \pi^- &\\
-\pi^- D_\mu \pi^+]\pi^0 \tau^3
+h_V^0 {a_6\over \sqrt{2}}\pi^0
D_\mu (\pi^+\pi^-) \tau^3\ \ \ .
\end{eqnarray}

The sum of all four possible sources, {\it i.e.}, Eq. (\ref{48}), (\ref{455}),
(\ref{46}) and (\ref{47}) yield
\begin{eqnarray}
\label{50}\nonumber
D_\mu (b_3 \pi^+\pi^-\pi^0 +b_4 \pi^0\pi^0\pi^0) \tau_3 &\\ \nonumber
-{1\over 6}h_V^0
[2\pi^+\pi^- D_\mu \pi^0
-\pi^0 D_\mu (\pi^+\pi^-)]\tau_3 &\\ \nonumber
+\sqrt{2}a_6 h_V^0 \pi^+\pi^-D_\mu\pi^0\tau_3 &\\
\nonumber
+{a_6\over \sqrt{2}}h_V^0 \pi^0 D_\mu (\pi^+\pi^-)\tau_3 &\\
+ i h_V^0{a_5\over \sqrt{2}}
[\pi^+D_\mu \pi^-
-\pi^- D_\mu \pi^+]\pi^0 \tau^3
\end{eqnarray}
In order for the transformation (\ref{eq:genunit}) to eliminate the $3\pi$
vector vertex, the sum in Eq. (\ref{50}) must vanish.
Note the first four lines and the last line of Eq. (\ref{50}) are,
respectively,
symmetric and anti-symmetric under the exchange $\pi^+\leftrightarrow \pi^-$.
The symmetric and anti-symmetric terms must vanish separately. The solution is
\begin{eqnarray}
\label{eq:vanish1}\nonumber
&b_3 =  -{2\over 3}h_V^0 \\ \nonumber
&b_4 =  0 \\ \nonumber
&a_5 =   0 \\
&a_6  = {1\over \sqrt{2}} \ \ \ .
\end{eqnarray}

Before considering the remaining $h_V^0$-terms, we observe that the
contributions from item (3) involves only
expressions involving the pion fields and $\tau_3, \tau_\pm$ multiplied by
real coefficients. The operator, ${\hat O}_2$, which contributes via items (1)
and (4), only appears in commutators. As a result, the three $\pi$ terms
involving
$a_{1-5}$ carry factors of $i$ and, thus, cannot cancel the contributions
in (3).
Consequently, we set $a_{1-5}=0$ in what follows.

Now we consider the terms linear in $h_V^0$ and $\tau_+$ (the argument for
$\tau_-$ is identical). The sum of these contributions is
\begin{eqnarray}
\label{conclusion}\nonumber
(b_6+ib_8) D_\mu (\pi^0\pi^0\pi^+) &\\ \nonumber
-{\sqrt{2}\over 6} h_V^0
\pi^0 (\pi^0 D_\mu \pi^+ -\pi^+ D_\mu \pi^0) &\\ \nonumber
-a_6h_V^0[ \pi^0\pi^0 D_\mu\pi^+
+ \pi^0 \pi^+ D_\mu\pi^0]&\\ \nonumber
+{\sqrt{2}\over 6} h_V^0
\pi^+ (\pi^+D_\mu \pi^--\pi^-D_\mu \pi^+) &\\
+ (b_5+i b_7) D_\mu (\pi^+ \pi^+\pi^-)
\end{eqnarray}
Clearly the last two lines (involving only charged $\pi$ fields) can never
cancel each
other.  The solution for the first three lines to vanish is
\begin{eqnarray}
\label{eq:vanish2}\nonumber
&a_6 = -{1\over \sqrt{2}} \\ \nonumber
&b_6 = -{\sqrt{2}\over 3} h_V^0\\
&b_8 =  0 \ \ \ .
\end{eqnarray}
Note that the requirements on $a_6$ in Eqs.
(\ref{eq:vanish1},\ref{eq:vanish2}) are
not consistent. Thus, it is not possible with the transformation
(\ref{eq:genunit})
to remove the $h_V^0$  $3\pi$ terms from the PV Lagrangian. Moreover,
as observed in Ref. \cite{kaplan}, Eq. (\ref{n1}) gives the most general
PV $\pi NN$ lagrangian up to one derivative of pion field. There
exist no additional PV vector $\pi NN$ contact interaction terms
which start off with three pions. Consequently, the $h_V^i$ cannot
not be absorbed as part of other LECs at three pion order.


{\bf Figure Captions}

\begin{center}
{\sf Figure 1.} {The relevant Feyman diagrams for PV $\pi^+$ photoproduction.
The circle filled with a cross is the PV vertex.}
\end{center}

\begin{center}
{\sf Figure 2.} {The t-channel $\rho$-meson exchange diagram used to estimate
the PV LEC ${\bar C}$.}
\end{center}

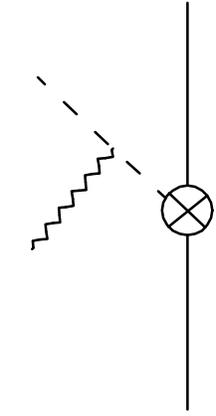
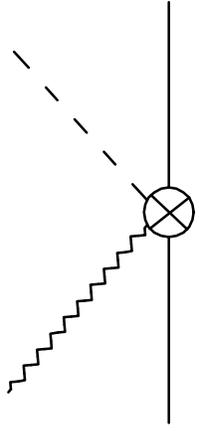
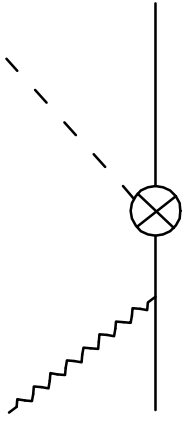
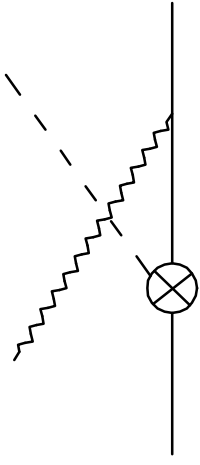
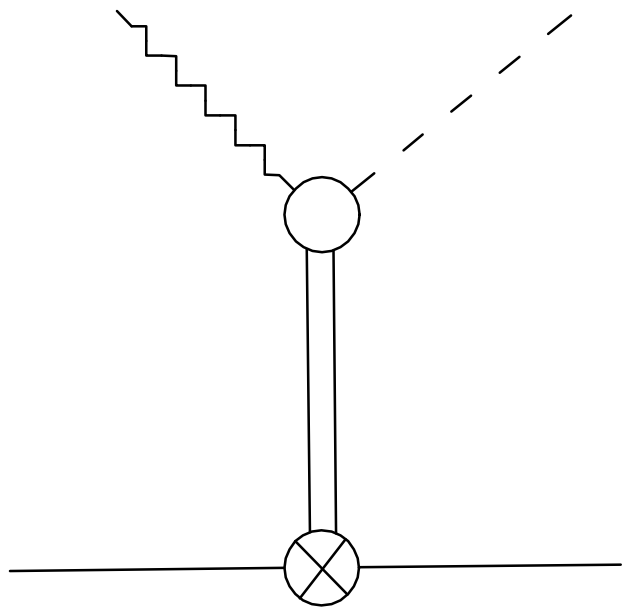
\end{document}